# Near-perfect Reachability of Variational Quantum Search with Depth-1 Ansatz

Junpeng Zhan[1]*

**Abstract**

**Grover's search algorithm is renowned for its dramatic speedup in solving many important scientific problems. The recently proposed Variational Quantum Search (VQS) algorithm has shown an exponential advantage over Grover's algorithm for up to 26 qubits. However, its advantage for larger numbers of qubits has not yet been proven. Here we show that the exponentially deep circuit required by Grover's algorithm can be replaced by a multi-controlled NOT gate together with either a single layer of $R_y$ gates or two layers of circuits consisting of Hadamard and NOT gates, which is valid for any number of qubits greater than five. We prove that the VQS, with a single layer of $R_y$ gates as its Ansatz, has near-perfect reachability in finding the good element of an arbitrarily large unstructured data set, and its reachability exponentially improves with the number of qubits, where the reachability is defined to quantify the ability of a given Ansatz to generate an optimal quantum state. Numerical studies further validate the excellent reachability of the VQS. Proving the near-perfect reachability of the VQS, with a depth-1 Ansatz, for any number of qubits completes an essential step in proving its exponential advantage over Grover's algorithm for any number of qubits, and the latter proving is significant as it means that the VQS can efficiently solve NP-complete problems.**

## 1. Introduction

Quantum algorithms can be broadly categorized into two types, depending on whether they are based on Grover's search algorithm or quantum Fourier transformation[1]. Grover's algorithm[2,3] provides a quadratic speedup in unstructured search and has numerous important applications[1]. However, the depth of the quantum circuit required in Grover's algorithm[2,3] grows exponentially with the number of qubits. To address this issue, a recently proposed algorithm called variational quantum search (VQS)[4] is capable of amplifying the total probability of the good element(s) to nearly 1 using a shallow circuit that grows linearly with the number of qubits, as verified up to 26 qubits due to the limitations of GPU memory. That is, the VQS shows an exponential advantage over Grover's algorithm in terms of circuit depth.

The VQS is a variational quantum algorithm (VQA)[5–7]. There is a lot of research on VQAs from different aspects, e.g., trainability[8–11], expressibility[12–14], reachability[15,16]. However, a similar analysis for the VQS has not been seen but is essential to verify the advantage of the VQS for any number of qubits, which is important as it determines whether the VQS can efficiently solve an NP-complete problem[17].

The reachability discusses the capability of a given Ansatz of VQA with parameters to represent a quantum state that minimizes a given objective function[8]. Reachability for QAOA[15,16,18] and variational Grover search[16,19] has been investigated. This paper focuses on the reachability of the VQS for unstructured search and proves that the exponentially deep circuit in Grover's algorithm can be replaced by either a single layer of $R_y(\theta)$ gates or a two-layer circuit consisting of Hadamard and $X$ gates. Furthermore, our numerical studies have verified the effectiveness of the VQS, with a single layer of $R_y(\theta)$ gates as its Ansatz, in solving the unstructured search problem. The rest of the paper consists of the Method, Result, and Conclusion sections.

[1] Department of Renewable Energy Engineering, Alfred University, Alfred, NY, USA. *E-mail: zhanj@alfred.edu

## 2. Method

The problem to be solved in this paper is to find the good element in an unstructured data set $\mathbb{D}$, which has only one good element and $(2^n - 1)$ bad elements. We use $n$ qubits and $n$ Hadamard gates to generate an equal superposition of all $2^n$ elements, as shown in the left-hand side of the leftmost dashed red line in Fig. 1a. In the rest of this section, we propose three methods to solve the problem: 1) a quantum circuit (see Fig. 1a,b) based on an HX layer, which is generated by Algorithm 1 (Section 2.1), 2) a quantum circuit (see Fig. 1c,d) based on an $R_y$ layer, which is generated by Algorithm 2 (Section 2.2), and 3) the VQS using the $R_y$ layer as the Ansatz (Section 2.4). Section 2.4 also discusses the reachability of the VQS. Section 2.3 clarifies the scalability of Algorithms 1 and 2.

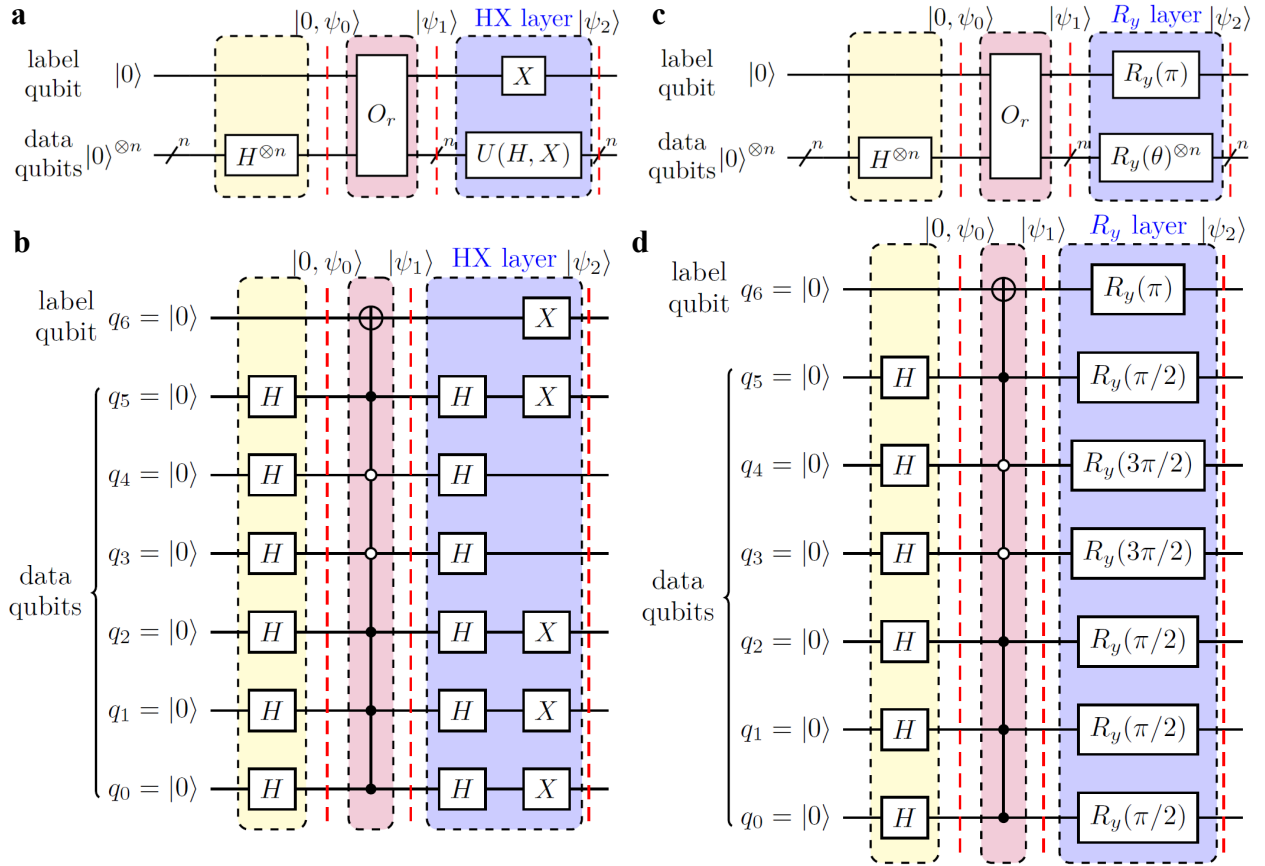

Fig. 1. The quantum circuit to generate data set $\mathbb{D}$ and amplify the probability of the only good element in it to nearly 1. **a,c,** the circuits in compact form for $n$-qubit data using an HX layer and an $R_y$ layer (the blue blocks in Fig. 1a and 1c), respectively, where the **HX layer** consists of Hadamard and $X$ gates and the $R_y$ **layer** consists of $R_y(\theta)$ gates. **b,d,** the detailed circuits for $n=6$

using the HX and $R_y$ layers, respectively. The yellow block, excluding the label qubit, generates a state that is an equal superposition of all elements (a single good element and $(2^n - 1)$ bad elements), i.e., all elements have the same initial probability. The red block (***Oracle***) provides label $|1\rangle$ at the label qubit to the good element and $|0\rangle$ to all bad elements. The blue block amplifies the probability of the good element to nearly 1. The label qubit is the highest one (the most significant one). In panel a, the $U(H, X)$ represents a circuit consisting of Hadamard and $X$ gates. In panels b,d, $n = 6$ and the index of the good element is 39 (its binary form is 100111).

## 2.1 Design the HX-layer-based Circuit

This section consists of four parts. The first part details an algorithm, Algorithm 1, which generates the HX layer (the blue block in Fig. 1a,b). The second part provides the vector forms of the three key quantum states in the HX-layer-based circuit, shown in Fig. 1a,b. The third part explains the design goal of the HX layer. The last part proves that Algorithm 1 can always achieve the design goal of the HX layer for any number of qubits.

**Algorithm 1.** Here we propose Algorithm 1 to construct the so-called HX layer, i.e., a two-layer circuit consisting of only Hadamard and $X$ gates. The HX layer together with the Oracle (the red blocks in Fig. 1) has the same function as Grover's search algorithm, i.e., amplify the probability of the good element to nearly 1.

**Algorithm 1 | Pseudo code for generating the HX layer (the blue block in Fig. 1a).**

---

**Input**: the number of qubits $n$ and the index of the good element in decimal form, $k, \forall k \in [0, 2^n - 1]$
**Output**: quantum gates in the HX layer (the blue block in Fig. 1a).
1    Convert $k$ into the binary form $b_{n-1}b_{n-2} \cdots b_2 b_1 b_0$.
2    Add an $X$ gate in the label qubit (the most significant qubit).
3    Let $m=n$
4    **while** $m \geq 1$
5        **if** $b_{m-1} = 1$
6           Add a Hadamard gate followed by an $X$ gate to qubit $q_{m-1}$.
7        **else**
8           Add a Hadamard gate to qubit $q_{m-1}$.
9        $m \leftarrow m-1$

---

**Vector Forms of Three Quantum States.** This sub-section details the vector forms of three quantum states, i.e., $|0, \psi_0\rangle$, $|\psi_1\rangle$, and $|\psi_2\rangle$, which are respectively indicated in the three dashed red lines in Fig. 1a.

The state, $|0, \psi_0\rangle$, can be written as $|0\rangle \otimes (|0\rangle + |1\rangle)^{\otimes n}$, which can be represented in the vector form:

$$|0, \psi_0\rangle = [\underbrace{\alpha_0^b, \alpha_1^b, \cdots, \alpha_{k-1}^b, \alpha_k^g, \alpha_{k+1}^b, \cdots, \alpha_{N-1}^b}_{\text{1st half: } N \text{ elements}}, \underbrace{0, 0, \cdots, 0}_{\text{2nd half: } N \text{ elements}}]^T \qquad (1)$$

where $N = 2^n$, super scripts b and g indicate bad and good elements, respectively, and subscripts, 0~$N$−1, represent the index of an element in the vector. Note that the index always counts from 0 in the rest of the paper. For example, $\alpha_k^g$ means the $k$th element is a good element.

The relationship between $|\psi_1\rangle$ and $|0, \psi_0\rangle$ can be represented as:

$$|\psi_1\rangle = O_r|0,\psi_0\rangle = [\underbrace{\alpha_0^b, \alpha_1^b, \cdots, \alpha_{k-1}^b, 0, \alpha_{k+1}^b, \cdots, \alpha_{N-1}^b}_{\text{1st half: } N \text{ elements}}, \underbrace{0,\cdots,0, \alpha_{N+k}^g, 0, \cdots, 0}_{\text{2nd half: } N \text{ elements}}]^T \quad (2)$$

where oracle $O_r$ is implemented as the $C^n(X)$, which is an $n$-qubit-controlled $X$ gate, as shown in Fig. 1b. As shown in Eq. (2), the $C^n(X)$ changes the index of the good element from $k$ to $N+k$.

The HX layer (the blue block in Fig. 1a) can be expressed as

$$X \otimes U(H,X) = \begin{bmatrix} 0 & U(H,X) \\ U(H,X) & 0 \end{bmatrix} \quad (3)$$

where $U(H,X)$ is an $N$ by $N$ matrix. For the convenience of expression, we call the matrix given in (3) the **HX-layer matrix**. Then the relationship between $|\psi_2\rangle$ and $|\psi_1\rangle$ can be represented as:

$$|\psi_2\rangle = \begin{bmatrix} 0 & U(H,X) \\ U(H,X) & 0 \end{bmatrix} |\psi_1\rangle$$

$$= \begin{bmatrix} 0 & U(H,X) \\ U(H,X) & 0 \end{bmatrix} [\underbrace{\alpha_0^b, \alpha_1^b, \cdots, \alpha_{k-1}^b, 0, \alpha_{k+1}^b, \cdots, \alpha_{N-1}^b}_{\text{1st half: } N \text{ elements}}, \underbrace{0,\cdots,0, \alpha_{N+k}^g, 0, \cdots, 0}_{\text{2nd half: } N \text{ elements}}]^T \quad (4)$$

For the convenience of analysis, we write $|\psi_2\rangle$ as a vector form:

$$|\psi_2\rangle = [\beta_0, \beta_1, \cdots, \beta_{N-1}, \beta_N, \cdots, \beta_{2N-1}]^T \quad (5)$$

where

$$\sum_{i=0}^{2N-1} |\beta_i|^2 = 1 \quad (6)$$

**Design Goal of the HX Layer.** The **goal** of designing the $U(H,X)$ is to let each element in its row $k$ be $1/\sqrt{N}$ such that $\beta_{N+k}$, the $(N+k)$th element of $|\psi_2\rangle$, can be calculated as:

$$\beta_{N+k} = \left(\sum_{i=0}^{k-1} \alpha_i^b + \sum_{i=k+1}^{N-1} \alpha_i^b\right)/\sqrt{N} \quad (7)$$

where the right-hand side is obtained from Eq. (4), i.e., multiply the $(N+k)$th row of the HX-layer matrix with the vector form of $|\psi_1\rangle$, and the integer $k \in [0, 2^n - 1]$.

From here on, we assume all elements have the same initial magnitude. That is, $\alpha_i^b = \alpha_k^g = \alpha_{N+k}^g = 1/\sqrt{N}$, $\forall i \in [0, N-1]$. Then, (7) can be reformed as:

$$\beta_{N+k} = (1/\sqrt{N})(N-1)/\sqrt{N} = 1 - 1/N = 1 - 1/2^n \quad (8)$$

The probability of obtaining the good element is equal to $\beta_{N+k}^2 = (1 - 1/2^n)^2$, which is equal to 0.25, 0.5625, 0.7656, 0.8789, 0.9386, 0.9690, 0.9844, 0.9922, and 0.9961 for $n$=1~9, respectively. That is, the probability of finding the good element is larger than 0.95 and 0.99 for $n$ being larger than 5 and 7, respectively.

**Position Index of the All-1 Row for the HX Layer.** In the rest of this sub-section, we answer the question of why the HX layer generated by Algorithm 1 can realize the design goal specified in the previous sub-section.

The result of tensor product $\begin{bmatrix} a_{n,0} \\ a_{n,1} \end{bmatrix} \otimes \begin{bmatrix} a_{n-1,0} \\ a_{n-1,1} \end{bmatrix} \otimes \cdots \otimes \begin{bmatrix} a_{2,0} \\ a_{2,1} \end{bmatrix} \otimes \begin{bmatrix} a_{1,0} \\ a_{1,1} \end{bmatrix} \otimes \begin{bmatrix} a_{0,0} \\ a_{0,1} \end{bmatrix}$ is a column vector with $2^{n+1}$ elements. The position index of the element $a_{n,i_n} a_{n-1,i_{n-1}} \cdots a_{2,i_2} a_{1,i_1} a_{0,i_0}$ in the column vector is equal to the decimal value of a binary form $i_n i_{n-1} \cdots i_2 i_1 i_0$, where $i_r \in \{0,1\}, \forall r \in [0,n]$. To better understand this, we provide two examples: the position index of the element $a_{3,0} a_{2,0} a_{1,0} a_{0,1}$ in the column vector associated with $\begin{bmatrix} a_{3,0} \\ a_{3,1} \end{bmatrix} \otimes \begin{bmatrix} a_{2,0} \\ a_{2,1} \end{bmatrix} \otimes \begin{bmatrix} a_{1,0} \\ a_{1,1} \end{bmatrix} \otimes \begin{bmatrix} a_{0,0} \\ a_{0,1} \end{bmatrix}$ is 1 (its binary form is 0001) and the position index of $a_{3,1} a_{2,0} a_{1,0} a_{0,1}$ in the same vector is 9 (its binary form is 1001).

We can express the tensor product $Y_{n-1} \otimes Y_{n-2} \otimes \cdots \otimes Y_2 \otimes Y_1 \otimes Y_0$ as $\frac{1}{\sqrt{2^n}} M$, where $M$ represents a $2^n$ by $2^n$ matrix and $Y_r, \forall r \in [0, n-1]$, represents either $XH$ or $H$. Note that $X = \begin{bmatrix} 0 & 1 \\ 1 & 0 \end{bmatrix}$, $H = \frac{1}{\sqrt{2}}\begin{bmatrix} 1 & 1 \\ 1 & -1 \end{bmatrix}$, $XH = \frac{1}{\sqrt{2}}\begin{bmatrix} 1 & -1 \\ 1 & 1 \end{bmatrix}$. It can be easily verified that $M$ has only one all-1 row (i.e., each element in the row is 1) while each of all the other rows consists of $-1$ and 1. The question then becomes, where is the all-1 row, which will be answered in the next paragraph.

This paragraph explains how to find the all-1 row for the tensor product given in the previous paragraph, and conversely, for any given number $k$, how to construct a quantum circuit such that row $k$ of the matrix associated with the circuit is an all-1 row. Given that each $Y_r$ has exactly one row of $[1 \ 1]/\sqrt{2}$, the all-1 row is generated from the tensor product of the $[1 \ 1]/\sqrt{2}$ row of each $Y_r$. In other words, if the $[1 \ -1]/\sqrt{2}$ row is involved, the corresponding tensor product result will not be all 1's. Let $j_r = 0$ represent that the $[1 \ 1]/\sqrt{2}$ is located in row 0 of $Y_r$ (i.e., $Y_r = H$) and $j_r = 1$ represent that the $[1 \ 1]/\sqrt{2}$ is located in row 1 of $Y_r$ (i.e., $Y_r = XH$). According to the paragraph before the previous one, we can know that the all-1 row is located at row $k$ where $k$ is the decimal value of its binary form $j_{n-1} j_{n-2} \cdots j_2 j_1 j_0$. Conversely, for any given $k \in [0, 2^n - 1]$, if we want to generate a matrix that represents the tensor product $Y_{n-1} \otimes Y_{n-2} \otimes \cdots \otimes Y_2 \otimes Y_1 \otimes Y_0$, such that each element in row $k$ of the matrix is 1 (ignoring the coefficient $1/\sqrt{2}$ of each $Y_r$), we can convert $k$ into its binary form $j_{n-1} j_{n-2} \cdots j_2 j_1 j_0$, set $Y_r$ to be $XH$ if $j_r = 1$, and set $Y_r$ to be $H$ if $j_r = 0$. This is exactly what Algorithm 1 does. To better understand this, we provide three examples for $k=5, 8$, and 39 in the next three paragraphs.

For $k=5$, its binary form is $j_2 j_1 j_0 = 101$. Then we have $Y_2 \otimes Y_1 \otimes Y_0$. By replacing $Y_2$ and $Y_0$ by $XH$ as $j_2 = j_0 = 1$, and replacing $Y_1$ by $H$ as $j_1 = 0$, we can obtain:

$$XH \otimes H \otimes XH = \frac{1}{\sqrt{2}}\begin{bmatrix} 1 & -1 \\ 1 & 1 \end{bmatrix} \otimes \frac{1}{\sqrt{2}}\begin{bmatrix} 1 & 1 \\ 1 & -1 \end{bmatrix} \otimes \frac{1}{\sqrt{2}}\begin{bmatrix} 1 & -1 \\ 1 & 1 \end{bmatrix}$$

$$= \frac{1}{2\sqrt{2}}\begin{bmatrix} 1 & -1 & 1 & -1 & -1 & 1 & -1 & 1 \\ 1 & 1 & 1 & 1 & -1 & -1 & -1 & -1 \\ 1 & -1 & -1 & 1 & -1 & 1 & 1 & -1 \\ 1 & 1 & -1 & -1 & -1 & -1 & 1 & 1 \\ 1 & -1 & 1 & -1 & 1 & -1 & 1 & -1 \\ 1 & 1 & 1 & 1 & 1 & 1 & 1 & 1 \\ 1 & -1 & -1 & 1 & 1 & -1 & -1 & 1 \\ 1 & 1 & -1 & -1 & 1 & 1 & -1 & -1 \end{bmatrix} \quad (9)$$

which shows that row 5 is the only all-1 row (ignoring the coefficient $\frac{1}{2\sqrt{2}}$). It is important to note that in this paper, the row index count always starts at 0, i.e., the first row is row 0.

For $k=8$, its binary form is 1000. Then we have

$$XH \otimes H \otimes H \otimes H = \frac{1}{\sqrt{2}}\begin{bmatrix} 1 & -1 \\ 1 & 1 \end{bmatrix} \otimes \frac{1}{\sqrt{2}}\begin{bmatrix} 1 & 1 \\ 1 & -1 \end{bmatrix} \otimes \frac{1}{\sqrt{2}}\begin{bmatrix} 1 & 1 \\ 1 & -1 \end{bmatrix} \otimes \frac{1}{\sqrt{2}}\begin{bmatrix} 1 & 1 \\ 1 & -1 \end{bmatrix} \quad (10)$$

We can easily verify that the tensor product shown in Eq. (10) has only one all-1 row (ignoring the coefficient ¼) located at row 8.

For $k=39$, its binary form is 100111. Then we can use $XH \otimes H \otimes H \otimes XH \otimes XH \otimes XH$ to obtain a matrix whose only all-1 row is located at row 39, as shown in Fig. 1b.

In summary, by using a $C^n(X)$ gate and an HX layer, we can amplify the probability of the good element from $1/2^n$ to a value larger than 0.95 and 0.99 for $n$ values greater than 5 and 7, respectively, where we already know the position index of the good element.

## 2.2 Design the $R_y$-layer-based Circuit

This section consists of four parts. The first part details an algorithm, Algorithm 2, to generate the $R_y$ layer (the blue block in Fig. 1c,d). The second part provides the vector form of the quantum state $|\psi_2\rangle$ in the $R_y$-layer-based circuit, shown in Fig. 1c,d. The third part explains the $R_y$ layer is equivalent to the HX layer in solving the problem of finding the good element.

**Algorithm 2.** Here we design an $R_y$ layer (the blue block in Fig. 1c) to replace the HX layer (the blue block in Fig. 1a), i.e., use a single layer of $R_y$ gates to replace the two layers consisting of $X$, $H$, and $XH$ gates. Given that $R_y(\theta) = \begin{bmatrix} \cos(\frac{\theta}{2}) & -\sin(\frac{\theta}{2}) \\ \sin(\frac{\theta}{2}) & \cos(\frac{\theta}{2}) \end{bmatrix}$, we have $R_y\left(\frac{\pi}{2}\right) = \frac{1}{\sqrt{2}}\begin{bmatrix} 1 & -1 \\ 1 & 1 \end{bmatrix}$ which is exactly equivalent to $XH$, $R_y(\pi) = \begin{bmatrix} 0 & -1 \\ 1 & 0 \end{bmatrix}$ which can be used to replace the $X$ gate despite the phase difference, and $R_y(3\pi/2) = \frac{1}{\sqrt{2}}\begin{bmatrix} -1 & -1 \\ 1 & -1 \end{bmatrix}$ which can be used to replace the $H$ gate despite the phase difference. The impact of the phase difference can be ignored with the reason given in the rest of this section. Algorithm 2 details the procedure of how to generate the $R_y$ layer.

**Algorithm 2 | Pseudo code for generating the $R_y$ layer (the blue block in Fig. 1c).**

**Input**: the number of qubits $n$ and the position index of the good element in decimal form, $k, \forall k \in [0, 2^n - 1]$
**Output**: quantum gates in the $R_y$ layer (the blue block in Fig. 1c).
1    Convert $k$ into the binary form $b_{n-1}b_{n-2}\cdots b_2b_1b_0$.
2    Add an $R_y(\pi)$ gate in the label qubit (the most significant qubit).
3    Let $m=n$
4    **while** $m \geq 1$
5        **if** $b_{m-1} = 1$
6            Add an $R_y(\pi/2)$ gate to qubit $q_{m-1}$.
7        **else**
8            Add an $R_y(3\pi/2)$ gate to qubit $q_{m-1}$.
9        $m \leftarrow m-1$

**Vector Forms of Quantum State $|\psi_2\rangle$.** The states $|0, \psi_0\rangle$ and $|\psi_1\rangle$ in Fig. 1c are the same as those in Fig. 1a. Therefore, only state $|\psi_2\rangle$ in Fig. 1c is provided here. The $R_y$ layer obtained from Algorithm 2, as shown in the blue blocks in Fig. 1c,d, can be expressed as

$$R_y(\pi) \otimes R_y(\theta)^{\otimes n} = \begin{bmatrix} 0 & -1 \\ 1 & 0 \end{bmatrix} \otimes R_y(\theta)^{\otimes n} = \begin{bmatrix} 0 & -R_y(\theta)^{\otimes n} \\ R_y(\theta)^{\otimes n} & 0 \end{bmatrix} \quad (11)$$

For the convenience of expression, we refer to the matrix given in Eq. (11) as the ***$R_y$-layer matrix***.

Like Eq. (4), based on Eq. (11), states $|\psi_2\rangle$ and $|\psi_1\rangle$ in Fig. 1c have the following relationship:

$$|\psi_2\rangle = \begin{bmatrix} 0 & -R_y(\theta)^{\otimes n} \\ R_y(\theta)^{\otimes n} & 0 \end{bmatrix} |\psi_1\rangle$$

$$= \begin{bmatrix} 0 & -R_y(\theta)^{\otimes n} \\ R_y(\theta)^{\otimes n} & 0 \end{bmatrix} [\underbrace{\alpha_0^b, \alpha_1^b, \cdots, \alpha_{k-1}^b, 0, \alpha_{k+1}^b, \cdots, \alpha_{N-1}^b}_{\text{1st half: } N \text{ elements}}, \underbrace{0, \cdots, 0, \alpha_{N+k}^g, 0, \cdots, 0}_{\text{2nd half: } N \text{ elements}}]^T \quad (12)$$

where $R_y(\theta)^{\otimes n}$ is an $N$ by $N$ matrix.

**Relation Between the $R_y$ Layer and the HX Layer.** This sub-section will explain that the $R_y$ layer is equivalent to the HX layer in solving the problem of finding the good element from an unstructured data set.

Similar to the analysis given in (4)-(7), the calculation of $\beta_{N+k}$, defined in Eq. (5), involves only the bottom left part, i.e., $R_y(\theta)^{\otimes n}$, of the $R_y$-layer matrix but not its top right part, i.e., $-R_y(\theta)^{\otimes n}$. Therefore, The phase difference between $R_y(\pi)$ and $X$ does not have any impact on the value of $\beta_{N+k}$. In other words, we can use $R_y(\pi)$ to replace $X$ gate without any impact on the probability of finding the good element.

We can write $R_y(\theta)^{\otimes n}$ as a tensor product $Y_{n-1} \otimes \cdots \otimes Y_2 \otimes Y_1 \otimes Y_0$, where $Y_r, \forall r \in [0, n-1]$, represents either $R_y(\pi/2)$ or $R_y(3\pi/2)$. Then $R_y(\theta)^{\otimes n}$ will have only one all-1 (all-negative-1) row if the number of $R_y(3\pi/2)$ gates is even (odd), where each element in the all-negative-1 row is $-1$, ignoring the coefficient $1/\sqrt{2^n}$. When the number of $R_y(3\pi/2)$ gates is even, the calculation given in (7) and (8) is also valid for Fig. 1c. When the number of $R_y(3\pi/2)$ gates is odd, each element in row $N+k$ is $-1$, ignoring the coefficient $1/\sqrt{2^n}$, then we have

$$\beta_{N+k} = -\left(\sum_{i=0}^{k-1} \alpha_i^b + \sum_{i=k+1}^{N-1} \alpha_i^b\right)/\sqrt{N} \tag{13}$$

$$\beta_{N+k} = -\left(1/\sqrt{N}\right)(N-1)/\sqrt{N} = -1 + 1/N = -1 + 1/2^n \tag{14}$$

Since the probability is the square of magnitude $\beta_{N+k}$, the magnitude of the good element given in (8) and that given in (14) result in the same probability. Therefore, no matter whether the number of $R_y(3\pi/2)$ gates is even or odd, the probability of finding the good element from the output of Fig. 1c, $|\psi_2\rangle$, is larger than 0.95 and 0.99 for $n$ values greater than 5 and 7, respectively, which is the same as the analysis for Fig. 1a given in the previous section (see the paragraph below Eq. (8)). Therefore, we can use the $R_y$ layer generated by Algorithm 2 to replace the HX layer generated by Algorithm 1.

## 2.3 Scalability of Algorithms 1 and 2

In the analysis for Figs. 1a and 1c above, there is neither restriction nor assumption associated with the number of qubits. That is, the analysis given in Sections 2.1 and 2.2 is valid for any number of qubits. Specifically, Algorithm 1 (Algorithm 2) each can generate a quantum circuit whose corresponding matrix form has only one all-1 (all-1 or all-negative-1) row located at row $k$, $k \in [0, 2^n - 1]$, for any value of $n$, where $n$ is the number of qubits.

In Algorithms 1 and 2, the location index, $k$, of the good element must be known in advance. This means that they are not useful for finding the good element in a data set if the location index is unknown. However, the analysis given above has proved that either the HX layer (generated by Algorithm 1) or the $R_y$ layer (generated by Algorithm 2), together with a $C^n(X)$ gate (the red and blue blocks in Fig. 1a,c), can amplify the probability of the good element from $1/2^n$ to nearly 1 if $n$ is larger than 5. Note that for the same task, Grover's search algorithm requires a quantum circuit whose depth increases exponentially with the number of qubits, which manifests the exponential advantage of the circuits generated by Algorithms 1 and 2, in terms of circuit depth.

## 2.4 Variational Quantum Search and its Reachability

This section briefs the VQS and then details the definition of reachability for the VQS. The VQS[4] is a VQA, which involves the interaction between classical and quantum computers. In the classical part, an optimizer is used to update the parameter $\boldsymbol{\theta}$ of the Ansatz based on the objective function $f(\boldsymbol{\theta})$:

$$f(\boldsymbol{\theta}) = -0.5\langle\psi_1|\psi_2\rangle + 0.5\langle\psi_1|Z \otimes I^{\otimes n}|\psi_2\rangle \tag{15}$$

where $\psi_1$ and $\psi_2$ are the states before and after the ansatz, respectively, as shown in Fig. 1a,c, $Z$ and $I$ are Pauli Z and identity matrix, respectively, and $n$ is the number of qubits. The first and second terms in the objective function can be respectively obtained from measuring two quantum circuits based on the Hadamard test, as detailed in paper[4]. Papar[4] has validated two types of shallow and effective Ansatzs for the VQS. However, the analysis given above implies that the VQS with only a single $R_y$ layer as the Ansatz, which is shallower than those two types of Ansatzs used in paper[4], is sufficient to find the only good element in an unknown data set. Note that unlike Algorithms 1 and 2, the VQS does not need to know the position index, $k$, of the good element beforehand. To quantify the capability of how best the Ansatz can generate a quantum state that minimizes a given objective function, we propose a definition of reachability for the VQS in the rest of this section.

The objective function given in Eq. (15) can be reformed as

$$f(\boldsymbol{\theta}) = \langle \psi_1 | O | \psi_2 \rangle \tag{16}$$

where

$$O = 0.5(-I^{\otimes n+1} + Z \otimes I^{\otimes n}) = \text{diag}([\underbrace{0,0,\cdots,0}_{N \text{ elements}}, \underbrace{-1,-1,\cdots,-1}_{N \text{ elements}}]) \tag{17}$$

where diag($w$) represents a square diagonal matrix with elements of vector $w$ on the main diagonal.

Then we can define a ***reachability***[15,16] for the VQS:

$$\mathfrak{R} = \frac{\min_{\theta}\langle \psi_1 | O | \phi(\theta) \rangle - \min_{|\varphi\rangle \in \mathcal{H}}\langle \psi_1 | O | \varphi \rangle}{-\min_{|\varphi\rangle \in \mathcal{H}}\langle \psi_1 | O | \varphi \rangle} \tag{18}$$

where $|\psi_1\rangle$ denotes the state before the ansatz (see Fig. 1c), the min$\theta$ term, used to refer to $\min_{\theta}\langle \psi_1 | O | \phi(\theta) \rangle$ for the convenience of expression, is the minimum over all reachable states generated by the Ansatz of the VQS, the $\phi(\theta)$ in the min$\theta$ term represents the $|\psi_2\rangle$ in Fig. 1c, and the minH term, used to refer to $\min_{|\varphi\rangle \in \mathcal{H}} \langle \psi_1 | O | \varphi \rangle$ for the convenience of expression, represents the minimum over all states $|\varphi\rangle$ of the Hilbert space $\mathcal{H}$.

This paragraph discusses the possible value range of the numerator in (18). Note that $|\psi_1\rangle$ has only a single non-zero element which is equal to $1/\sqrt{N}$ in this paper and located in its second half and that $\langle x|O|y\rangle$ is equal to the tensor product of the second half of $-|x\rangle$ and the second half of $|y\rangle$, where the second half of a state represents the second half elements of the vector form of the state. We can then know that, in the numerator in (18), the value range of $\langle \psi_1|O|\phi(\theta)\rangle$ is a subset of $[-1/\sqrt{N}, 1/\sqrt{N}]$ and the value range of $\langle \psi_1|O|\varphi\rangle$ is $[-1/\sqrt{N}, 1/\sqrt{N}]$. Therefore, the range of $\langle \psi_1|O|\phi(\theta)\rangle - \langle \psi_1|O|\varphi\rangle$ is a subset of $[-2/\sqrt{N}, 2/\sqrt{N}]$, which is an extremely small range when $n$ is large as both the upper and lower bounds converge exponentially towards zero.

Given that a well-defined metric should not consistently reside within a minuscule range around zero, using the numerator in (18) as a measure of reachability is not appropriate as it is extremely small when the number of qubits is large and converges exponentially towards zero as the increase of the number of qubits, regardless of the value of $\theta$. To address the issue, we use the minH term in the denominator as explained in the next paragraph.

According to the analysis given in paper[4], the minimum value of the minH term, i.e., $-1/\sqrt{N}$, is achieved when $|\varphi\rangle$ is equal to the computational state $|N + k\rangle$, which is represented by a vector with a single element of 1 at index $N+k$ and all other elements being 0's, where $N = 2^n$. Given that dividing the range $[-2/\sqrt{N}, 2/\sqrt{N}]$ by the minH term results in a range of $[-2, 2]$, we use the minH term as the denominator such that the range of $\mathfrak{R}$ defined in (18) becomes a subset of $[0, 2]$ instead of the extremely small range mentioned above. A range of $[0, 2]$ is more suitable for the reachability metric, as it is not confined to a minuscule range around zero, and this is the reason why we introduce the denominator in the definition of reachability.

Now we explain why we add the absolute sign into each term, as shown in Eq. (19). As per Eqs. (8) and (14), the value of $\beta_{N+k}$ being equal to either $1 - 1/N$ or $-1 + 1/N$ results in the same probability of finding the good element, i.e., both solutions are equally good, where $N = 2^n$. However, putting these two values of $\beta_{N+k}$ into $|\psi_2\rangle$, which replaces $|\phi(\theta)\rangle$, results in different values of $\langle\psi_1|O|\phi(\theta)\rangle$ and thereby leads to different values of reachability, but they should correspond to the same reachability. To address this issue, we use absolute terms, i.e., $|\langle\psi_1|O|\phi(\theta)\rangle|$ and $|\langle\psi_1|O|\varphi\rangle|$, to replace $\langle\psi_1|O|\phi(\theta)\rangle$ and $\langle\psi_1|O|\varphi\rangle$, respectively, such that two amplitudes with different signs lead to the same reachability.

This paragraph discusses a better definition of reachability, as given in Eq. (19). Note that $\max_{|\varphi\rangle\in\mathcal{H}}|\langle\psi_1|O|\varphi\rangle|$ reaches its maximum when $\langle\psi_1|O|\varphi\rangle$ equals $1/\sqrt{N}$ or $-1/\sqrt{N}$, corresponding to finding the good element with a probability of 1. On the other hand, $\min_{|\varphi\rangle\in\mathcal{H}}|\langle\psi_1|O|\varphi\rangle|$ reaches its minimum when $\langle\psi_1|O|\varphi\rangle$ equals 0, corresponding to finding the good element with a probability of 0. Thus, we use 'max' rather than 'min' when using the absolute terms such that we can find the good element with a high probability. According to the discussion given in the previous paragraph, two amplitudes with different signs lead to different values of reachability if we use (18) as the definition of reachability, which could be addressed if we use Eq. (19). Therefore, we recommend using Eq. (19) as the definition of ***reachability*** instead of (18).

$$\mathfrak{R} = \frac{\max_{\theta}|\langle\psi_1|O|\phi(\theta)\rangle| - \max_{|\varphi\rangle\in\mathcal{H}}|\langle\psi_1|O|\varphi\rangle|}{-\max_{|\varphi\rangle\in\mathcal{H}}|\langle\psi_1|O|\varphi\rangle|} \tag{19}$$

The smaller the value of $\mathfrak{R}$, the better the reachability. When $\mathfrak{R} = 0$, the VQS has perfect reachability. The use of denominator and absolute terms distinguishes our definition of reachability from that presented in papers[15,16].

For the circuit depicted in Fig. 1c where the $R_y$ layer is generated by Algorithm 2, the ($N+k$)th element of the output state $|\psi_2\rangle$ is either equal to $1 - 1/N$ or $-1 + 1/N$, as shown in Eqs. (8) and (14), respectively. If we replace the $\max_{\theta}|\langle\psi_1|O|\phi(\theta)\rangle|$ with $|\langle\psi_1|O|\psi_2\rangle|$ in Eq. (19), we have

$$\frac{|\langle\psi_1|O|\psi_2\rangle| - \max_{|\varphi\rangle\in\mathcal{H}}|\langle\psi_1|O|\varphi\rangle|}{-\max_{|\varphi\rangle\in\mathcal{H}}|\langle\psi_1|O|\varphi\rangle|} = \frac{\frac{1-1/N}{\sqrt{N}} - \frac{1}{\sqrt{N}}}{-\frac{1}{\sqrt{N}}} = \frac{1}{N} = \frac{1}{2^n} \tag{20}$$

As $|\psi_2\rangle$ is generated by Algorithm 2, it is the $|\phi(\theta)\rangle$ with the $\theta$ specified in Algorithm 2. Then we know $\max_{\theta}|\langle\psi_1|O|\phi(\theta)\rangle| \geq |\langle\psi_1|O|\psi_2\rangle|$. Therefore, we have

$$\mathfrak{R} = \frac{\max_{\theta}|\langle\psi_1|O|\phi(\theta)\rangle| - \max_{|\varphi\rangle\in\mathcal{H}}|\langle\psi_1|O|\varphi\rangle|}{-\max_{|\varphi\rangle\in\mathcal{H}}|\langle\psi_1|O|\varphi\rangle|} \leq \frac{|\langle\psi_1|O|\psi_2\rangle| - \max_{|\varphi\rangle\in\mathcal{H}}|\langle\psi_1|O|\varphi\rangle|}{-\max_{|\varphi\rangle\in\mathcal{H}}|\langle\psi_1|O|\varphi\rangle|} = \frac{1}{2^n} \quad (21)$$

where the two sides of the ≤ come from Eqs. (19) and (20), respectively. Eq. (20) provides a specific circuit, i.e., using the $R_y$ layer generated by Algorithm2 as the Ansatz of the VQS, whose reachability value is equal to $1/2^n$, which is an extremely small value when $n$ is large and converges exponentially to zero as the increase of $n$. That is, this circuit already achieves near-perfect reachability.

In conclusion, the VQS has at least near-perfect reachability, and the more qubits, the better the reachability of the VQS.

## 3. Result

The results provided in this paper are obtained from quantum simulators using Pennylane's devices[20]. Results related to 2-, 8-, and 14-qubit (20- and 26-qubit) input states are obtained using Pennylane's default.qubit (lightning.gpu) device on an Intel i5-6500 CPU (A40x4 48-GB GPU). The initial values of $\boldsymbol{\theta}$ in the Ansatz are randomly sampled from a uniform distribution between 0 and $2\pi$. Same as paper[4], two termination criteria for the iterative process are used in the VQS. The first one is that the number of iterations reaches a given number (set to 300 in this paper). The second one is that a small-change event occurs consecutively for a given number of times (set to 5 in this paper), where the small-change event is defined as: the absolute value of the relative change of objective functions in two consecutive iterations is smaller than a given small value (set to $1 \times 10^{-4}$ in this paper). When one of the criteria is met, whichever comes first, the iterative process of VQS terminates.

Here we numerically verify that the VQS with only one layer of $R_y(\theta)$ gates can efficiently find the good element from an unstructured data set without knowing the position index, $k$, of the good element beforehand, which is related to the trainability of the VQS. We apply the VQS, where the Ansatz is an $R_y$ layer, to find the good element from 2-, 8-, 14-, 20-, and 26-qubit data sets.

The results are shown in Fig. 2, which show that the VQS indeed can find the good element as the amplified probability is very close to 1 for most runs out of 100 runs for $n$=8, 14, 20, and 26, and that in some runs (22, 16, 16, and 16 for the four cases) out of 100 runs, the amplified probability is close to 0. Compared to the results shown in paper[4], although the VQS using the $R_y$ layer in the Ansatz is less stable than the VQS using the two types of Ansatzs shown in paper[4], the former can still amplify the probability to nearly 1 in most runs, excluding the 2-qubit case. The relatively poor performance of the 2-qubit case is not a concern, but actually further validates our analysis as explained below. When $n$ equals 2 and if we use the $R_y$ layer as the Ansatz of the VQS, according to Eq. (8), the probability of finding the good element is $(1 - 1/2^2)^2 = 0.5625$, which is roughly in the middle of the box result for the 2-qubit case (the leftmost one in Fig. 2a). In other words, the results in Fig. 2a validate the conclusion obtained from the reachability analysis, i.e., the more the number of qubits, the better the reachability of the VQS.

The importance of using VQS with the $R_y$ layer is that it has been proved above that the $R_y$ layer, together with the $C^n(X)$, can amplify the probability of the good element from $1/2^n$ to nearly 1 for any number of qubits being larger than 5. That is, the reachability of the VQS using the $R_y$ layer is guaranteed for any number of qubits being larger than 5, which means we do can scale the VQS to any large number of qubits while keeping the circuit depth to be 2 (i.e., one $R_y$ layer and one $C^n(X)$ layer), which is an exponential advantage over Grover's algorithm.

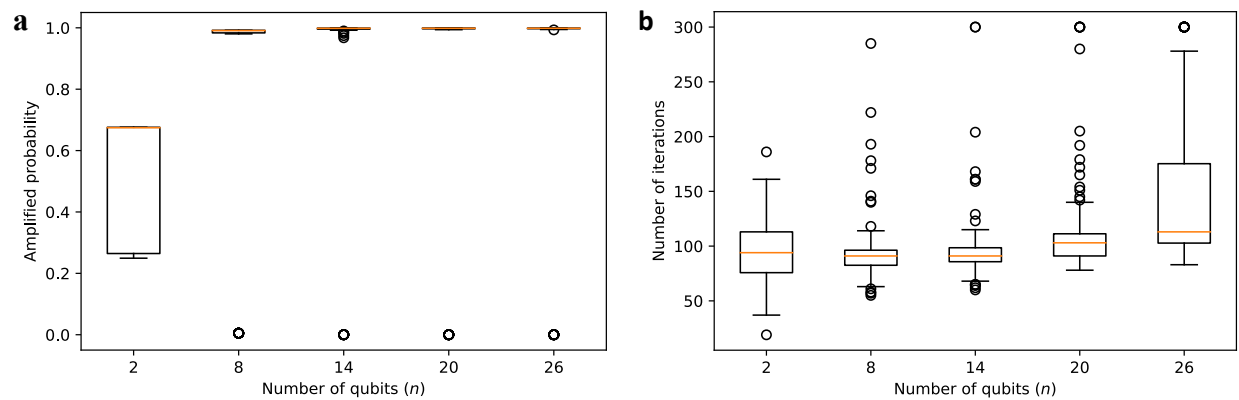

fig. 2. **Box plot results from 100 runs of VQS** using the $R_y$ layer as the Ansatz for an *n*-qubit input state. **a,** the amplified probability of good element. **b**, the number of iterations used when a termination criterion is met.

## 4. Conclusion

We have proposed two algorithms to construct a depth-2 (the HX layer) and a depth-1 (the $R_y$ layer) circuits, both of which can replace the exponentially deep circuit required by Grover's algorithm, if the position index of the good element is known beforehand. We have proved that the VQS, with the $R_y$ layer as the Ansatz, has near-perfect reachability for any number of qubits greater than five and its reachability exponentially improves with the number of qubits, which has been further verified by numerical experiments. In these experiments, we use the VQS, with the $R_y$ layer as Ansatz, to search for the only good solution in unstructured data sets, represented by 8, 14, 20, and 26 qubits, and successfully find the good element with a probability of nearly 1 in 78 to 84 out of 100 independent runs. The experiments also indicate the good trainability of the VQS for up to 26 qubits. We will further investigate the trainability of the VQS for more qubits in the future.

## Acknowledgements

This research was partially supported by the NSF ERI program, under award number 2138702. This work used the Delta system at the National Center for Supercomputing Applications through allocation CIS220136 from the Advanced Cyberinfrastructure Coordination Ecosystem: Services & Support (ACCESS) program, which is supported by National Science Foundation grants #2138259, #2138286, #2138307, #2137603, and #2138296. We acknowledge the use of IBM Quantum services for this work. The views expressed are those of the authors, and do not reflect the official policy or position of IBM or the IBM Quantum team.